\documentclass[prb,superscriptaddress,twocolumn,showpacs]{revtex4-1}
\usepackage{graphicx}
\usepackage{lipsum}
\usepackage{hyperref}

\begin{document}

\title{Influence of heat flow directions on Nernst effects in Py/Pt bilayers}

\author{D.\ Meier}
\email{dmeier@physik.uni-bielefeld.de}
\homepage{www.spinelectronics.de}
\author{D.\ Reinhardt}
\affiliation{Thin Films and Physics of Nanostructures, Department of Physics, Bielefeld University, D-33501 Germany}
\author{M.\ Schmid}
\author{C.H.\ Back}
\affiliation{Institute of Experimental and Applied Physics, University of Regensburg, D-93040, Germany}
\author{J.-M.\ Schmalhorst}
\author{T.\ Kuschel}
\author{G.\ Reiss}
\affiliation{Thin Films and Physics of Nanostructures, Department of Physics, Bielefeld University, D-33501 Germany}

\date{November 7, 2013}

\begin{abstract}

We investigated the voltages obtained in a thin Pt strip on a Permalloy film which was subject to in-plane temperature gradients and magnetic fields. The voltages detected by thin W-tips or bond wires showed a purely symmetric effect with respect to the external magnetic field which can be fully explained by the planar Nernst effect (PNE). To verify the influence of the contacts measurements in vacuum and atmosphere were compared and gave similar results. We explain that a slightly in-plane tilted temperature gradient only shifts the field direction dependence but does not cancel out the observed effects. Additionally, the anomalous Nernst effect (ANE) could be induced by using thick Au-tips which generated a heat current perpendicular to the sample plane. The effect can be manipulated by varying the temperature of the Au-tips. These measurements are discussed concerning their relevance in transverse spin Seebeck effect measurements.

\end{abstract}


\maketitle

\section{Introduction}

\begin{figure}[t!]%
\includegraphics[width=\linewidth]{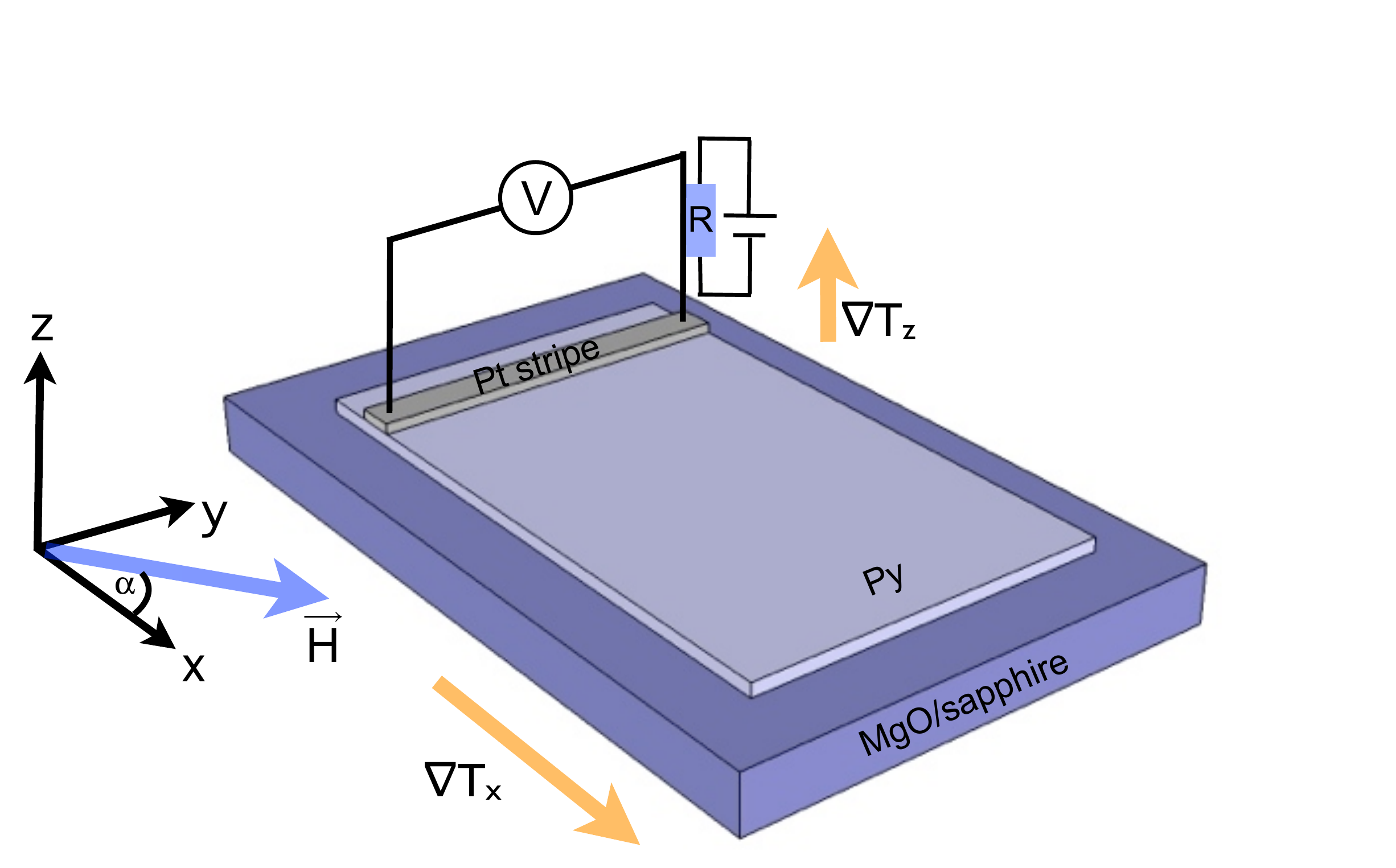}%
\caption{Sample geometry. A 10\,nm Pt strip on top of a 20\,nm Py film on a MgO or sapphire substrate. An in-plane temperature gradient \(\nabla T_x\) is applied as well as an external magnetic field \(\vec{H}\) with an angle \(\alpha\) with respect to the x-direction. Additionally, an out-of-plane temperature gradient \(\nabla T_z\) can occur and can be influenced when a voltage is applied at a resistor \(R\) which heats up one of the contact tips. The resistor is electrical insulated from the tip. The generated voltage is measured at the ends of the Pt strip.}%
\label{sample}%
\end{figure}

Thermoelectric and spincaloritronic effects have become an emerging field since the observation of the spin Seebeck effect (SSE) in Py thin films.\cite{Uchida:2008cc} An in-plane temperature gradient \(\nabla T_x\) generates a spin current which can be detected in a Pt strip on the Py film. The direction of the spin polarization is given by the magnetization in the magnetic layer. This spin current is measured as an electromotive force \(\vec{E}_{\text{ISHE}}\) of the inverse spin Hall effect (ISHE) \cite{Saitoh:2006kk} in the Pt strip given by: 
\begin{equation}\label{inversespinhalleffect}
\vec{E}_{\text{ISHE}}=D_{\text{SHE}}\vec{J}_S \times \vec{\sigma}.
\end{equation}
\(\vec{E}_{\text{ISHE}}\) is perpendicular to the spin current \(\vec{J}_S\) and the spin polarization vector \(\vec{\sigma}\) which is aligned to an external magnetic field in the sample plane. \(D_{\text{SHE}}\) describes the magnitude of the spin Hall effect (SHE)\cite{Hirsch:1999wr,Valenzuela:2006cs} and therefore relates the ISHE to the SSE.\cite{Uchida:2008cc} A theoretical explanation of the SSE was given by Xiao et al.\cite{Xiao:2010iy} The main characteristics of the voltage signal are the proportionality to the temperature difference between the ends of the sample and the sign change of the voltage by changing the Pt position from the hot to the cold end of the Py film (for \(\alpha\)\,=\,0\(^\circ\) in Fig. \ref{sample}). Furthermore, a sign change of the voltage can be observed by changing the direction of the external magnetic field due to the cross product of the ISHE. The latter leads to a switching of the voltage with the coercive field of the Py. Furthermore, a cos(\(\alpha\)) symmetry can be observed using an in-plane magnetic field rotation with angle \(\alpha\) between the magnetic field and the x-direction.\cite{Uchida:2008cc}

The SSE was also observed in a broad range of materials like ferromagnetic semiconductors\cite{Jaworski:2010dy}, Heusler compounds\cite{Bosu:2011bw} and ferrimagnetic insulators\cite{Uchida:2010ei}. The detected spin current has to propagate perpendicularly to the Py film into the Pt and therefore perpendicular to the in-plane temperature gradient. This configuration is called transverse SSE (TSSE). An alternative configuration for SSE measurements is the so called longitudinal SSE (LSSE). The LSSE has been observed in ferrimagnetic insulators like Y\(_3\)Fe\(_5\)O\(_{12}\) (YIG) \cite{Uchida:2010jb} and other garnet ferrites \cite{Uchida:2013dz}. Furthermore, the LSSE was shown for ferrites like (Mn,Zn)Fe\(_2\)O\(_4\), for ferrimagnetic semiconducting NiFe\(_2\)O\(_4\) \cite{Meier:2013dz} and for Fe\(_3\)O\(_4\) (magnetite) below the Verwey transition \cite{Ramos:2013ab}. There, the temperature gradient is applied perpendicular to the sample (\(\nabla T_z\)) which generates a spin current into a Pt film on top. Now the spin current propagates parallel to the temperature gradient and an electromotive force can be measured at the Pt due to the ISHE. In the longitudinal configuration parasitic effects like the anomalous Nernst effect (ANE) can occur for metallic systems and semiconductors.\cite{Meier:2013dz}

The ANE is the anomalous counterpart of the conventional Nernst effect\cite{nernst:1886ab} and describes the generation of a transverse voltage perpendicular to both a temperature gradient and the magnetization, given by: 
\begin{equation}\label{anomalousnernsteffect}
\vec{E}_{\text{ANE}}= N_{\text{ANE}} \nabla T \times \vec{M}.
\end{equation}
\(\vec{E}_{\text{ANE}}\) is a cross product of the temperature gradient \(\nabla T\) and the magnetization \(\vec{M}\) and is proportional to a coefficient N\(_{\text{ANE}}\). 

In TSSE measurements one can also observe parasitic effects such as the ANE. Therefore, one has to avoid perpendicular temperature gradients to prevent a voltage contribution due to the ANE induced in the magnetic and conductive Py film \cite{Huang:2011cd,Bosu:2011bw}. For the ANE the voltage is proportional to the temperature gradient and a cos(\(\alpha\)) symmetry can be observed using an in-plane magnetic field rotation. Here for large magnetic fields the magnetization \(\vec{M}\) is parallel to the external field \(\vec{H}\) and, therefore, the angle \(\alpha\) is equal to an angle \(\varphi\) between \(\vec{M}\) and \(\nabla T_x\). The sign of the voltage is changing with the magnetic field direction. Nevertheless, a sign change of the ANE can not been observed if the Pt strip is moved from the hot to the cold end of the sample, unless an out-of-plane temperature gradient is changing its direction. For parasitic out-of-plane temperature gradients which change direction from hot to cold side of the sample one cannot distinguish between the TSSE and the ANE from the measured voltage. Furthermore, both effects are antisymmetric with respect to the external magnetic field.

In ferromagnetic conductors an additional effect can be generated by an in-plane temperature gradient, this is the planar Nernst effect (PNE).\cite{Pu:2006cra,Ky:1966ur} The PNE is the thermally generated counterpart to the anisotropic magnetoresistance (AMR) also called anisotropic magnetothermopower (AMTEP).\cite{Schmid:2013m} When a temperature gradient \(\nabla T_x\) is applied along the x-direction (Fig.\,\ref{sample}) a voltage in y-direction can be observed which depends on the direction of the magnetization vector \(\vec{M}\) of the ferromagnet:
\begin{equation}\label{planarnernsteffect}
V_y \propto |M|^2 \text{sin}(\varphi)\text{cos}(\varphi) |\nabla T_x| \propto M_x \cdot M_y |\nabla T_x|
\end{equation}
where \(\varphi\) is the angle between \(\nabla T_x\) and \(\vec{M}\). The maximum of this effect can be found for \(\varphi\)\,=\,45\(^\circ\) and \(\varphi\)\,=\,225\(^\circ\), the minimum for \(\varphi\)\,=\,135\(^\circ\) and \(\varphi\)\,=\,315\(^\circ\) and the effect should vanish for \(\varphi\)\,=\,0\(^\circ\), \(\varphi\)\,=\,90\(^\circ\), \(\varphi\)\,=\,180\(^\circ\) and \(\varphi\)\,=\,270\(^\circ\). The PNE is a symmetric effect with respect to the external magnetic field. Therefore, the PNE can be separated from the antisymmetric SSE and the antisymmetric ANE. There have been several reports which discuss the contribution of PNE and even ANE in measurements of the TSSE in Py films with and without a Pt strip. \cite{Huang:2011cd,Avery:2012bj,Schmid:2013m,Yin:2013by}

In this work we present measurements with a uniform applied in-plane temperature gradient and a detailed discussion of the angle dependencies of the observed voltages. Furthermore, we show results with an intentionally induced out-of-plane temperature gradient through the contact tips on the Pt strip. The results are discussed in view of the SSE, ANE and PNE.

\section{Experimental}

We prepared 3\,\(\times\)\,5\,mm\(^2\) Py films on 10\,x\,10\,mm\(^2\) MgO and sapphire substrates by e-beam evaporation and sputtering. Due to the deposition process a small uniaxial magnetic anisotropy occurred \cite{chikazumi:1956ab,Dillinger:1935ia}. Different samples were prepared where in some cases the substrate was cleaned by Ar\(^+\) etching before the deposition of the Py film to investigate the influence of different interface treatments. Finally, on one side of the sample a 10\,nm thin and 100\,\(\mu\)m wide Pt strip was deposited by dc magnetron sputtering with an Ar pressure of 1.5\,x\,10\(^{-3}\)\,mbar (Fig.\,\ref{sample}). The preparation was done in-situ at a base pressure of 3\,x\,10\(^{-9}\)\,mbar.

The measurements were performed in two similar setups. In general both consist of two copper blocks with copper plates to clamp the sample with the (Py free) substrate edge between the copper blocks. On each copper block a thermocouple is placed to obtain the temperature difference \(\Delta T_x\) between the ends of the sample. One setup (S1) is located in vacuum to avoid thermal convection. The Pt strip was contacted via 25\,\(\mu\)m thin Au bonding wires which were glued to copper cables with silver paste. The second setup is working in atmosphere and the Pt strip was connected by a micro probe system with Au and W-tips with a thermal conductivity of 318\,W/m\(\cdot\)K and 173\,W/m\(\cdot\)K respectively.\footnote{data sheet of the manufacturer} For the voltage measurement a Keithley 2182A Nanovoltmeter was used in both setups.

\section{Results}

\begin{figure}[t!]%
\includegraphics[width=\linewidth]{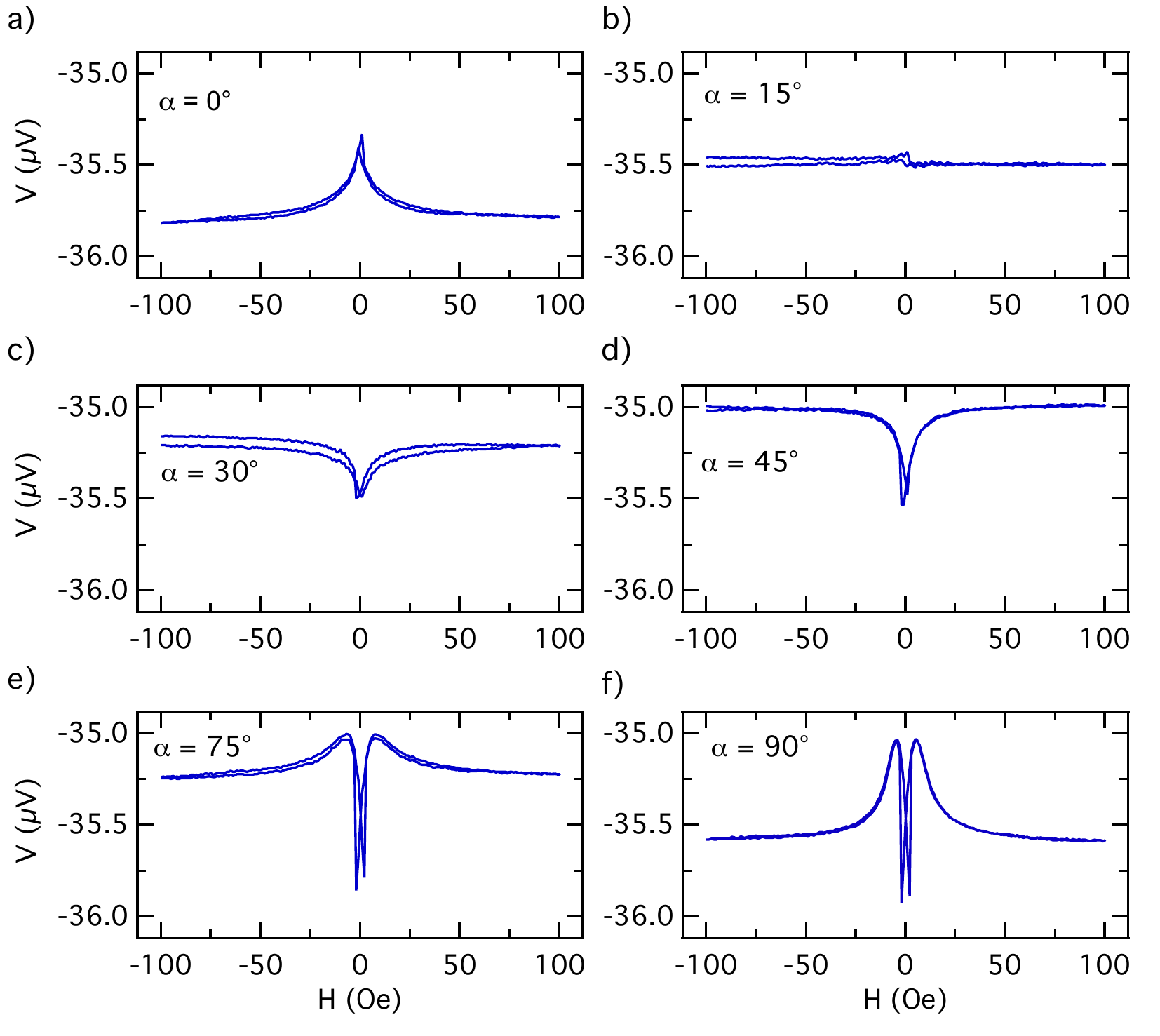}%
\caption{Measurements of the voltage signal in a Pt strip on the hot side of a Py film on an MgO substrate. The temperature difference between the ends of the Py film were 35\,K and the lower temperature was stabilized at 298\,K. The external magnetic field was varied in the in-plane direction with an angle \(\alpha\) in regards to the x-direction. The Pt strip was connected with Au bonding wires and the measurements were performed in vacuum.}%
\label{DM120112a_Furnace_1}%
\end{figure}%

In Fig.\,\ref{DM120112a_Furnace_1} the voltage \(V\) measured in S1 at the ends of the Pt strip is shown as a function of the external magnetic field \(\vec{H}\). A temperature difference \(\Delta T_x\) of about 35\,K was applied at a base (lower) temperature of 298\,K. The external magnetic field was applied in-plane along different directions with the angle \(\alpha\) with respect to the x-direction (Fig.\,\ref{sample}). For \(\alpha\)\,=\,\(0^\circ\) a symmetric curve with respect to the external field was obtained (Fig.\,\ref{DM120112a_Furnace_1}a)). For large magnetic fields the voltage saturates and at around \(0\)\,Oe the voltage shows maxima represented by two peaks close to each other, one for each branch of the curve. For \(\alpha\)\,=\,\(15^\circ\) the difference between the saturated and the maximum values nearly vanishes and the curve shows a constant voltage of about \(\text{-35.5}\)\,\(\mu V\) (Fig.\,\ref{DM120112a_Furnace_1}b)). For \(\alpha\)\,=\,\(30^\circ\) and \(\alpha\)\,=\,\(45^\circ\) the saturated voltage is larger than the voltage around \(0\)\,Oe. This minimum voltage can again be found around the value of about -35.5\,\(\mu V\). For \(\alpha\)\,=\,\(75^\circ\) and \(\alpha\)\,=\,\(90^\circ\) two peaks around \(0\)\,Oe (minimum voltage) and two maximum values for small magnetic fields occur which are different from the saturated voltage for large magnetic fields. It is remarkable that the crossing point at \(H\)\,=\,\(0\)\,Oe is around the offset value of about -35.5\,\(\mu V\) which is the conventional thermopower induced in the contact wires. The offset value varies for different wiring conditions and could also be observed by Schmid et al.\cite{Schmid:2013m}

The appearance of a symmetric and the absence of an antisymmetric effect lead to the assumption the PNE is the origin of the signal of Fig.\,\ref{DM120112a_Furnace_1} and that the SSE and the ANE can be excluded as an explanation.\cite{Schmid:2013m}

\begin{figure}[t!]%
\includegraphics[width=\linewidth]{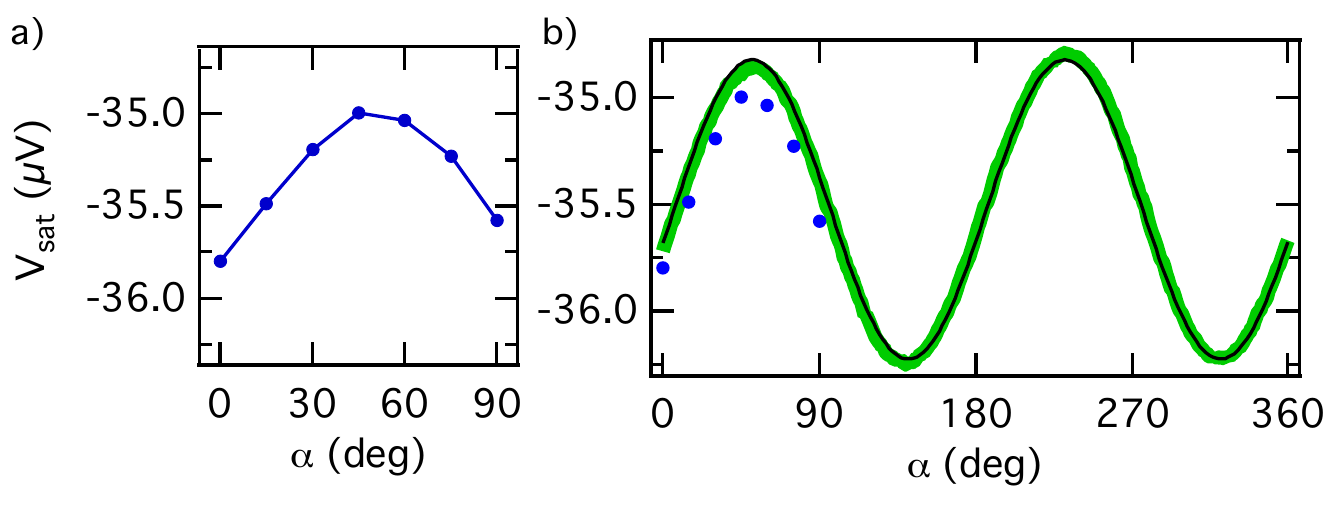}%
\caption{Angle dependent measurement of the voltage signal with \(\Delta T_x\)\,=\,35\,K at room temperature in vacuum. The Pt strip was contacted with Au bonding wires. \(\alpha\) is the angle between the x-direction and the external magnetic field \(\vec{H}\). a) the saturated voltage signal \(V_{\text{sat}}\) for large magnetic fields determined from the curves in Fig.\,\ref{DM120112a_Furnace_1} shown as a function of \(\alpha\). b) the saturated voltage \(V_{\text{sat}}\) shown as a function of a rotating magnetic field with \(|\vec{H}|\)\,=\,100\,Oe (bold line) compared with a fit sin function (thin line).}%
\label{DM120112a_Furnace_angle_1}%
\end{figure}%

\begin{figure}[t!]%
\includegraphics[width=2.8in]{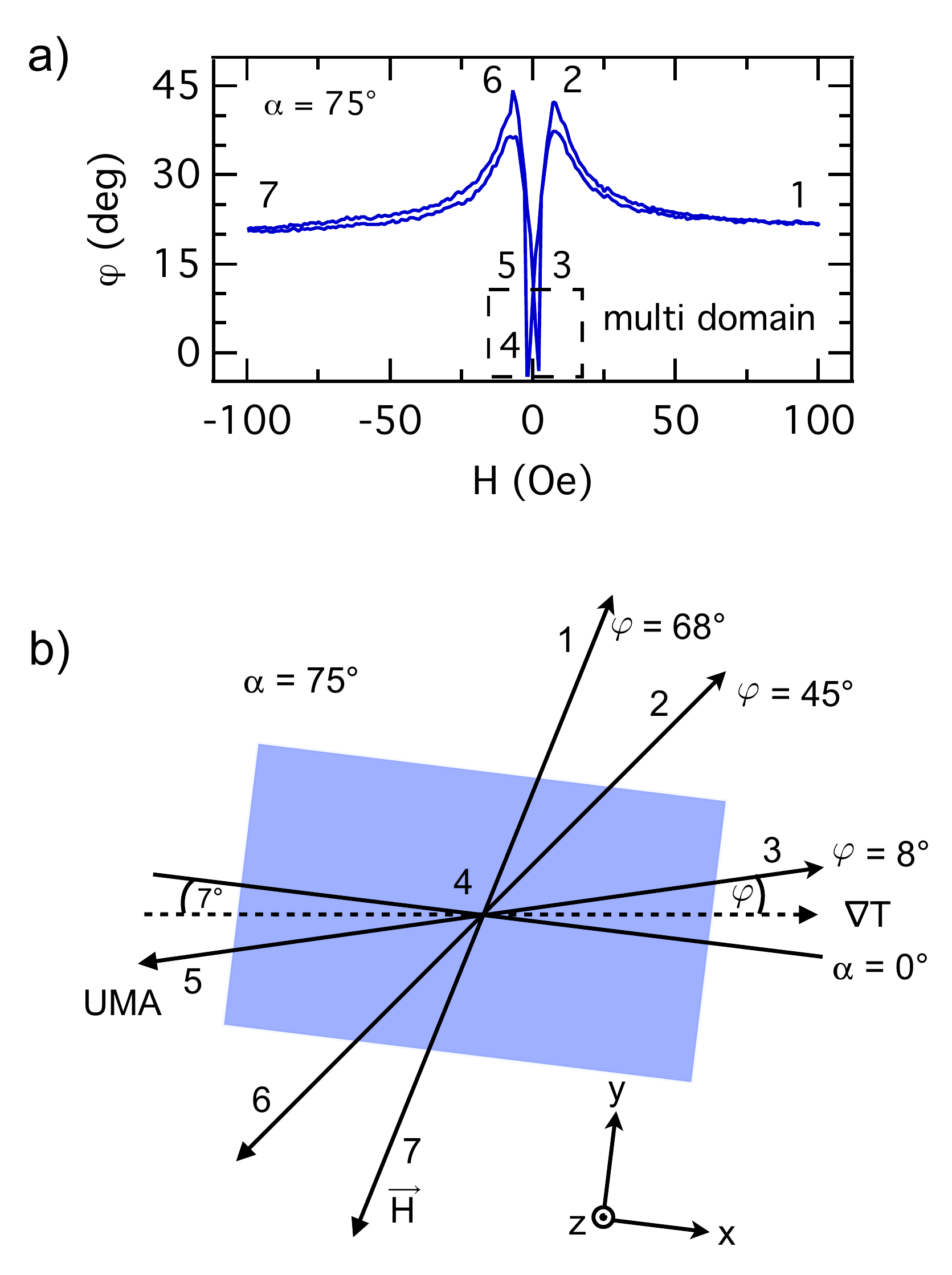}%
\caption{a) The angle \(\varphi\) of the magnetization vector \(\vec{M}\) of the Py film with respect to the temperature gradient \(\nabla T_x\) as a function of the external magnetic field H applied at an angle \(\alpha\)\,=\,75\(^\circ\) with respect to the long sample axis (\(\alpha\)\,=\,0\(^\circ\)).  The applied temperature gradient \(\nabla T_x\) shows a shift of 7\(^\circ\). The range of \(\varphi\) is limited from 0\(^\circ\) to 45\(^\circ\) due to the \(\frac{1}{2}\)arcsin calculation. In the multi domain state the magnetization \(M\) is decreasing without changing the value \(\varphi\) (dashed box). b) The reversal process of the magnetization vector \(\vec{M}\) of the Py film during the variation of the external magnetic field \(\vec{H}\) with an angle \(\alpha\)\,=\,75\(^\circ\). When \(|\vec{H}|\) is decreased \(\vec{M}\) rotates coherently from no. 1 via no. 2 into the magnetic easy direction (no. 3) due to a uniaxial magnetic anisotropy. In magnetic remanence (no. 4) the total magnetization is zero due to the multi domain state during the magnetic switching. When \(\vec{H}\) is increased again the mono domain state is reached (no. 5). After that, \(\vec{M}\) is rotating coherently via no. 6 to no. 7 (\(\vec{M}\parallel \vec{H}\)).}%
\label{angle_calc_uma_pne}%
\end{figure}

In Fig.\,\ref{DM120112a_Furnace_angle_1}\,a) the saturated voltage \(V_{\text{sat}}\) for large magnetic fields is shown as a function of the angle \(\alpha\) for the results of Fig.\,\ref{DM120112a_Furnace_1}. For the estimation of \(V_{\text{sat}}\) the data between -100\,Oe to -90\,Oe and 90\,Oe to 100\,Oe were averaged for each \(\alpha\). \(V_{\text{sat}}\) shows a maximum for \(\alpha\)\,=\,\(45^\circ\), where the maximum of the PNE is expected (Eq. \ref{planarnernsteffect}). The angle dependence was confirmed with a rotation of the magnetic field between \(\alpha\)\,=\,\(0^\circ\) and \(\alpha\)\,=\,\(360^\circ\) using \(|\vec{H}|\,=\,100\,\text{Oe}\) to maintain the sample in saturation. Fig.\,\ref{DM120112a_Furnace_angle_1}\,b) shows that the observed sin(\(\alpha\))\(\cdot\)cos(\(\alpha\))\,\(\propto\)\,sin(2\(\alpha\)) symmetry was achieved which confirms Eq.\,\ref{planarnernsteffect} for magnetic saturation (\(\varphi\)\,\(\approx\)\,\(\alpha\)). The calculated fit function gives an amplitude of 700\,nV and a phase shift of \(\alpha_0\)\,=\,-7\(^\circ\) for the sin(2\(\cdot(\alpha+\alpha_0\))) term. Therefore, the temperature gradient \(\nabla T_x\) is off the x-direction by the angle of -7\(^\circ\) (Fig.\,\ref{angle_calc_uma_pne}\,b)). This leads to \(\varphi\)\,=\,\(\alpha\)\,-\,7\(^\circ\) in magnetic saturation. Schmid et al.\cite{Schmid:2013m} explained in their supplements that a small transverse temperature gradient \(\nabla T_y\) gives a phase shift in the \(V_{\text{sat}}\) signal when the magnetic field is rotated in the sample plane.

Due to the constant line in Fig.\,\ref{DM120112a_Furnace_1}\,b) for \(\alpha\)\,=\,15\(^\circ\) one can conclude that the Py film has a magnetic easy axis along \(\alpha\)\,=\,15\(^\circ\) induced by a uniaxial magnetic anisotropy (UMA). When the field is applied parallel to the easy axis the magnetization of the Py film points along the same direction. The magnetization vector \(\vec{M}\) switches from \(\alpha\)\,=\,15\(^\circ\) to \(\alpha\)\,=\,345\(^\circ\) when the external magnetic field changes its sign from positive to negative field values, but the product of \(M_x \cdot M_y\) keeps its sign and also the measured voltage remains constant. 

Recently, an angle dependence of the PNE in thin Py films without a Pt strip has been shown.\cite{Yin:2013by} Here, we want to give a detailed description for the magnetization reversal process. Therefore, the results in Fig.\,\ref{DM120112a_Furnace_1} can be converted and interpretated in terms of the angle \(\varphi\) between \(\vec{M}\) and \(\nabla T_x\) as a function of the external magnetic field \(\vec{H}\) as shown for \(\alpha\)\,=\,75\(^\circ\) (Fig.\,\ref{angle_calc_uma_pne}). The maximum voltage is achieved for \(\varphi\)\,=\,45\(^\circ\) due to the \(M_x\)\,\(\cdot\)\,\(M_y\) product (Eq. \ref{planarnernsteffect}). The voltage can be normalized \(V_{\text{norm}}\) such that the maximum voltage corresponds to the value 1 for \(\varphi\)\,=\,45\(^\circ\) (=sin(2\(\cdot\)45\(^\circ\))) and the saturated value (at H\,=\,\(\pm\)100\,Oe) for \(\varphi\)\,=\,\(\alpha\)\,-\,7\(^\circ\) is related to sin(2\(\cdot\)(\(\alpha\)-7\(^\circ\))). In case of \(\alpha\)\,=\,75\(^\circ\) the saturation value is therefore related to sin(2\(\cdot\)(75\(^\circ\)\,-\,7\(^\circ\)))\,\(\approx\)\,0,7. Now, one can calculate the angle \(\varphi\)\,=\,\(\frac{1}{2}\)arcsin(\(V_{\text{norm}}\)) and plot it against \(|\vec{H}|\) (Fig.\,\ref{angle_calc_uma_pne}\,a)). Regarding the range of the \(\frac{1}{2}\)arcsin function only angles between 0\(^\circ\) and 45\(^\circ\) can be accessed. Nevertheless, one can use this graph to reconstruct the complete reversal process of the magnetization.

For \(\alpha\)\,=\,\(75^\circ\) the magnetization reversal is sketched in Fig.\,\ref{angle_calc_uma_pne}\,b). The angle \(\varphi\) corresponds to the same angle in Fig.\,\ref{angle_calc_uma_pne}\,a). The temperature gradient is shifted by -7\(^\circ\) as shown in the magnetic field rotation measurement. For large magnetic fields the saturated voltage is smaller than the maximum value of the PNE curve. When the external magnetic field is decreased the magnetization vector rotates coherently into the magnetic easy axes (no. 1-3 in Fig.\,\ref{angle_calc_uma_pne}\,b)). During this rotation the magnetization vector crosses \(\varphi\)\,=\,\(45^\circ\) (no. 2 in Fig.\,\ref{angle_calc_uma_pne}\,b)) and the voltage signal reaches its maximum (no. 2 in Fig.\,\ref{angle_calc_uma_pne} a)). In magnetic remanence the magnetic easy axis is reached and the voltage of about -35.5\,\(\mu V\) is obtained (no. 3 in Fig.\,\ref{angle_calc_uma_pne} a) and Fig.\,\ref{angle_calc_uma_pne}\,b)). Afterwards, a switching of the magnetization can be observed. Here, the magnetic mono domain state changes to a magnetic multi domain state and the voltage signal decreases due to the reduction of the total magnetic moment (no. 4 in Fig.\,\ref{angle_calc_uma_pne}\,b)). This multi domain state is similar to a coherent rotation of the magnetization vector from \(\varphi\)\,=\,8\(^\circ\) to 0\(^\circ\) (dashed box in Fig.\,\ref{angle_calc_uma_pne}\,a)). When the external magnetic field is increased in the opposite direction the mono domain state is reached again with \(\vec{M}\) pointing into the direction of \(\varphi\)\,=\,352\(^\circ\) (no. 5 in Fig.\,\ref{angle_calc_uma_pne}\,b)) which is similar to 8\(^\circ\) in Fig.\,\ref{angle_calc_uma_pne}\,a). Afterwards, the magnetization vector crosses the angle \(\varphi\)\,=\,315\(^\circ\) and the PNE gets another maximum (no. 6 in Fig.\,\ref{angle_calc_uma_pne}\,b)). When the external magnetic field is large enough \(\vec{M}\) is parallel to \(\vec{H}\) and the voltage signal becomes saturated again (no. 7 in Fig.\,\ref{angle_calc_uma_pne}\,b)). For \(\alpha\)\,=\,90\(^\circ\) in Fig.\,\ref{DM120112a_Furnace_1} the difference between the voltage for large external magnetic fields and the maximum value at lower fields is even more distinctive.

These experiments thus showed, that only symmetric voltages with respect to the magnetic field were measured, which are in full agreement with the PNE. Therefore, the SSE and ANE can be excluded. Nevertheless, this leads to the conclusion that in the given setup only an in-plane temperature gradient was applied and parasitic perpendicular temperature gradients can be neglected. This behavior was found for Py/Pt samples with and without Ar\(^+\) cleaned substrate and Py film.

In the work of Xiao et al.\cite{Xiao:2010iy} it was supposed that the thermally driven spin current is generated by the difference between the magnon temperature in the Py film and the electron temperature in the Pt strip. However, recent works show that there is no significant temperature difference to generate a measurable spin current in this configuration. This was shown using numerical simulations by Schreier et al.\cite{Schreier:2013tm} and experimentally observed with Brilliouin light scattering by Agrawal et al.\cite{PhysRevLett.111.107204} for the ferrimagnetic insulator YIG. Furthermore, the work of Schmid et al.\cite{Schmid:2013m} reveals that there is an upper limit of a possible TSSE contribution of about 20\,nV.

\begin{figure}[h!]%
\includegraphics[width=\linewidth]{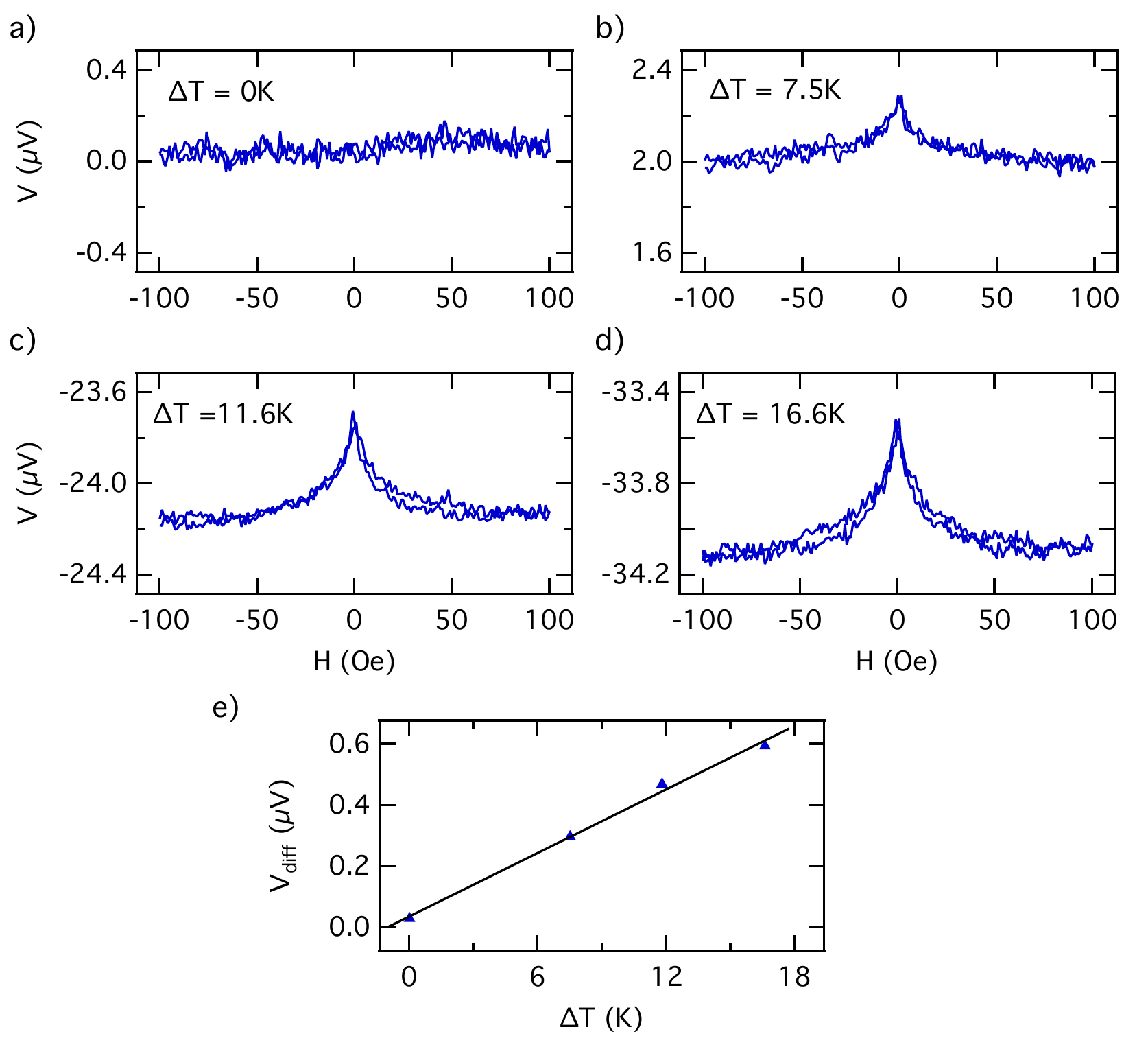}%
\caption{a) - d) Temperature dependence of \(V(|\vec{H}|)\) in atmosphere at room temperature. The Pt strip was connected with W-tips and the external magnetic field H was applied parallel to \(\nabla T_x\). e) The difference \(V_{\text{diff}}\)\,=\,\(V_{\text{max}}\)\,-\,\(V_{\text{sat}}\) between the saturated voltage signal \(V_{\text{sat}}\) for large external magnetic fields \(\vec{H}\) and the maximum voltage \(V_{\text{max}}\) around \(H\,=\,0\,\text{Oe}\) as a function of the temperature difference \(\Delta T_x\).}%
\label{DM120112a_GMR_1}%
\end{figure}%

For comparison the same sample was measured in another setup (S2) contacted with a micro probing system in atmosphere. In this setup the external magnetic field \(\vec{H}\) was applied parallel to an in-plane temperature gradient \(\nabla T_x\). Fig.\,\ref{DM120112a_GMR_1} shows the voltage at the Pt strip as a function of the external magnetic field. The results can also be explained by the PNE and compared with Fig.\,\ref{DM120112a_Furnace_1} for \(\alpha\)\,=\,0\(^\circ\). Parasitic out-of-plane temperature gradients can be neglected again. It can also be verified that the change of the PNE (difference \(V_{\text{diff}}\)\,=\,\(V_{\text{max}}\)\,-\,\(V_{\text{sat}}\) between saturated voltage \(V_{\text{sat}}\) for large external magnetic fields and maximum voltage \(V_{\text{max}}\)) is proportional to \(|\nabla T_x|\) as given by Eq. \ref{planarnernsteffect} (Fig.\,\ref{DM120112a_GMR_1}\,e)). The extrapolation of the fitted linear function in Fig.\,\ref{DM120112a_GMR_1}\,e) leads to a voltage signal of 1.2\,\(\mu V\) for \(\Delta T_x\)\,=\,35K and is comparable to the magnitude in Fig.\,\ref{DM120112a_Furnace_1}\,a), but due to different sample lengths between the copper blocks during the sample mounting in both setups the absolute value of the PNE magnitude differs. This verifies the equal functionality for both setups and shows identical results by using bonding wires in vacuum and W-tips in atmosphere. 

\begin{figure}[h!]%
\includegraphics[width=\linewidth]{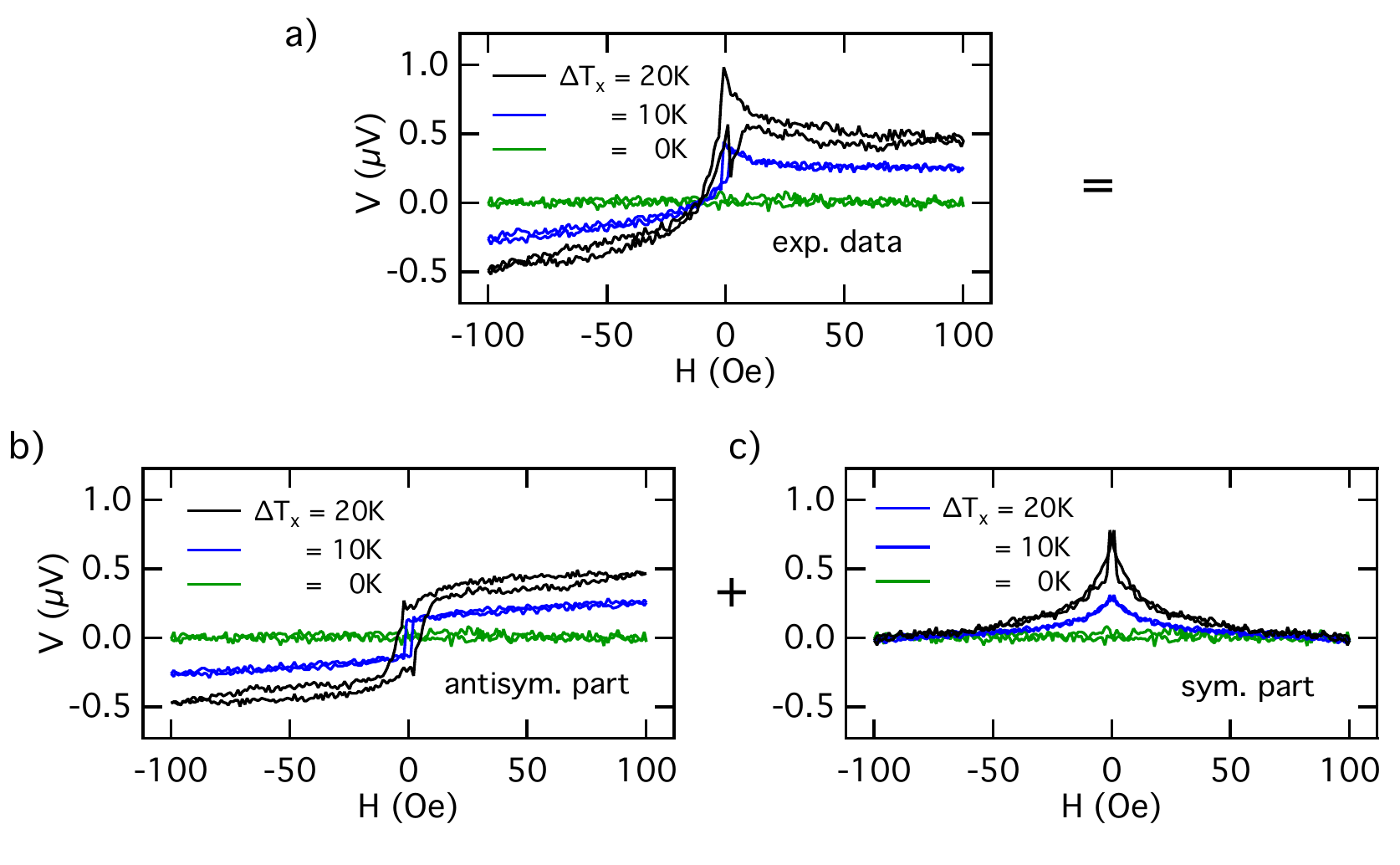}%
\caption{a) Measurements with three different in-plane temperature gradients at room temperature. The sample was connected with two Au-tips with a diameter of about 1\,mm. The obtained data were separated mathematically into an antisymmetric part (b) and a symmetric part (c).}%
\label{DM120112a_2AuTipsA}%
\end{figure}%

In the next step the measurement was modified by using the same samples but Au-tips with a diameter of about 0.5\,mm at the tip and 1\,mm at the tail instead of W-tips with diameters of about 10\,\(\mu\)m at the tip and 500\,\(\mu\)m at the tail and a larger thermal conductivity to analyze the influence of this different probe parameters. Fig.\,\ref{DM120112a_2AuTipsA}\,a) shows the measured voltage as a function of the external magnetic field for \(\Delta T_x\)\,=\,0\,K, 10\,K and 20\,K. The results contain an antisymmetric part in addition to the already known symmetric part. These parts were separated mathematically by calculating the sum and the difference of the two branches. For a better comparison an offset voltage was subtracted and the curves were centered around zero. The magnitude of the antisymmetric effect as well as the change of the symmetric effect is proportional to \(\Delta T_x\). The symmetric part can again be assigned to the PNE. The antisymmetric part can only be induced by the use of Au-tips instead of W-tips. Since the Au-tips are thicker than the W-tips this leads to the assumption that a heat current perpendicular to the sample plane flows through the Au-tips. Therefore, the antisymmetric part can be explained by the ANE due to this induced perpendicular temperature gradient. A TSSE can be excluded, since it was not observable using the W-tips.

\begin{figure}[h!]%
\includegraphics[width=\linewidth]{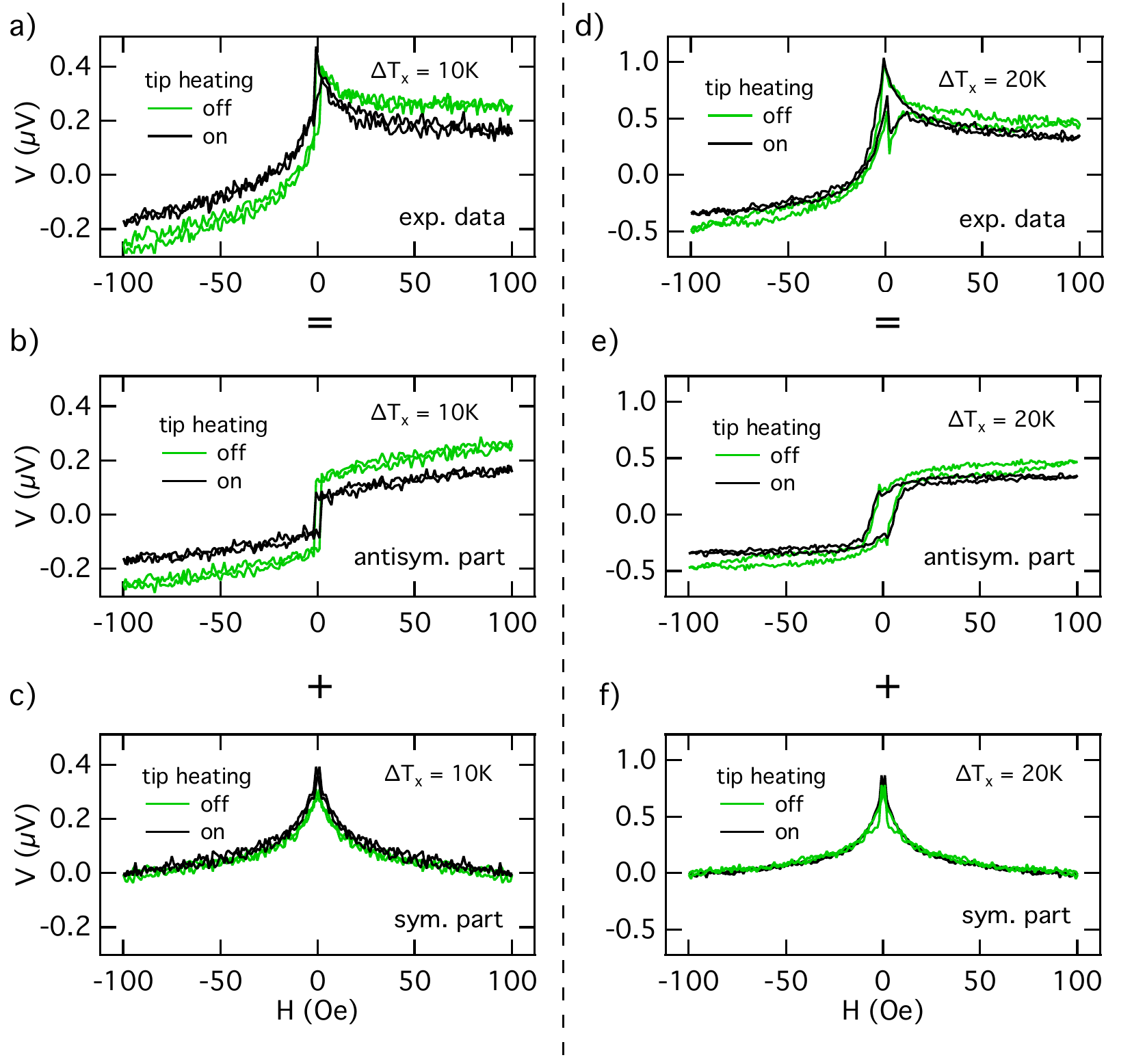}%
\caption{a), d) Measurements with two different in-plane temperature gradients at room temperature. The sample was contacted with a conventional Au tip and a heatable Au tip with a diameter of about 1\,mm respectively. Curves were obtained with and without tip heating. Each curve was separated mathematically into an antisymmetric part b),e) and a symmetric part c),f).}%
\label{DM120112a_2AuTipsB}%
\end{figure}%

\begin{figure}[h!]%
\includegraphics[width=2in]{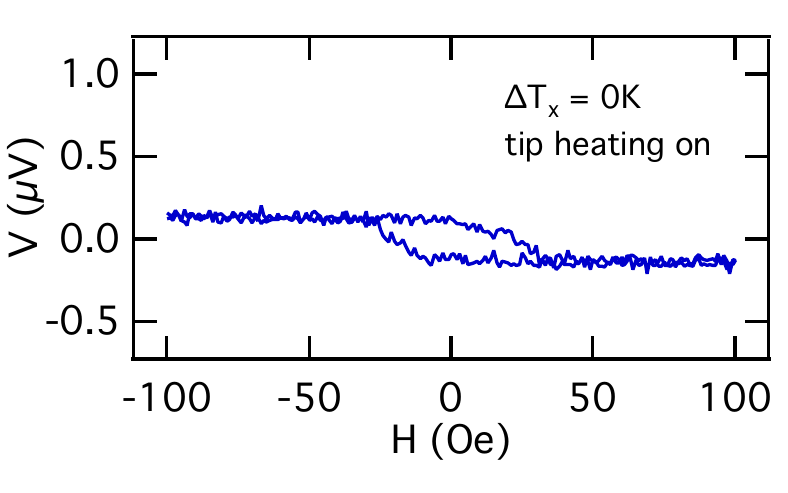}%
\caption{\(V\) as a function of \(H\); only the Au-tip was heated and the temperature difference between the sample ends was \(\Delta T\)\,=\,0\,K.}%
\label{DM120112a_2AuTips_0K_10V}%
\end{figure}%

The induced perpendicular heat current through the Au-tips can be manipulated by heating the tips. Therefore, one of the Au-tips was heatable by a fixed 1.5\,k\(\Omega\) resistor (applied voltage of about 10\,V and 15\,V). Fig.\,\ref{DM120112a_2AuTipsB} shows the measurements for \(\Delta T_x\)\,=\,10\,K and 20\,K with and without a heated Au tip (10\,V at the resistor for \(\Delta T_x\)\,=\,10\,K and 15\,V for \(\Delta T_x\)\,=\,20\,K). For the discussion, it is important to know, that the Pt strip and hence the Au-tips where positioned at the hot side on the Py film. The obtained curves were separated into the symmetric and antisymmetric parts to compare the effect of the heated tip. It is shown that the antisymmetric part which presents the ANE is decreased. The symmetric part which presents the PNE is not affected and the curves are equal for heated and not heated tip. The difference between the saturated voltage signals for \(\Delta T_x\)\,=\,10\,K and 10\,V at the resistor is about 90\,nV and for \(\Delta T_x\)\,=\,20\,K and 15\,V at the resistor about 125\,nV.

The influence of the heated Au-tip without an in-plane temperature gradient is shown in Fig.\,\ref{DM120112a_2AuTips_0K_10V}. For 10\,V at the resistor and \(\Delta T_x\)\,=\,0\,K at the sample ends. The measured voltage shows only an antisymmetric part with hysteretic behavior. It has a magnitude of about 250\,nV (the difference between the saturated voltage signals for large magnetic fields). The different sign of the measurement with heated Au-tips without \(\nabla T_x\) is in good agreement with the measurements in Fig.\,\ref{DM120112a_2AuTipsB}. The heat current of the heated Au-tip compensates the heat current which flows from the heated sample into the Au-tip. Therefore, the antisymmetric part in Fig.\,\ref{DM120112a_2AuTipsB} gets smaller by heating the Au-tip. The observed magnitude of 250\,\(\mu\)V can be compared with the work of Yin et al. \cite{Yin:2013by} regarding the missing Pt strip. A 30\,nm Py film was fully covered by a Cu-heater to apply an out-of-plane temperature gradient. They found a magnitude of nearly 250\,\(\mu\)V for \(\Delta T\)\,=\,2\,K. This emphasizes the influence of this kind of sensitive measurements by the contact method. 

We would like to emphasize that the obtained results do not show any TSSE signal within the given accuracy of the measurement. This is in accordance with previous work of i.e. Huang et al. \cite{Huang:2011cd}, Avery et al. \cite{Avery:2012bj} and Schmid et al. \cite{Schmid:2013m} The reasons why the TSSE does not appear in our measurements could be an insufficient spin mixing conductance due to the interface between the Py/Pt layers which is important for the spin current to enter the Pt film. Although different surface treatments were performed on different samples the spin mixing conductance could be still below a certain limit to generate a detectable voltage due to the ISHE.

The induced heat flow in out-of-plane direction due to thicker Au tips were only observed on the hot side of the Py film. On the cold side, the magnitude of the ANE was vanishing low due to comparable temperatures between the sample and the Au tips. Therefore, a sign change between the hot and the cold side of the Py film can appear for the ANE, which can be misinterpreted as a TSSE. This could happen when the Au tips are heated by a hot atmosphere in the setup chamber which is hotter than the cold side of the sample, but colder than the hot side. For this case, one has opposite out-of-plane temperature gradients for hot and cold side.

\section{Conclusion}

In conclusion, we investigated Py/Pt bilayers with both an in-plane temperature gradient and magnetic field. The \(V_{\text{ISHE}}\) voltages show a symmetric characteristic with respect to \(\vec{H}\) and can be fully described by the PNE. The reversal process of the magnetization vector of the Py film could be explained after performing angle dependent measurements and was completely reconstructed by the measured curves. Even a slightly in-plane tilted temperature gradient can be included in the interpretation and does not cancel out the observed effects. The results could be reproduced with a bonding wire connected sample in vacuum and with the same sample connected with thin W-tips in atmosphere. An antisymmetric effect explained by the ANE was induced by thicker Au-tips. An explanation for the ANE is a heat current generated through the tips perpendicular to the sample plane. The magnitude of this antisymmetric effect could be manipulated by heating the tips. These results give an extensive impression about parasitic effects and measurement methods for the investigation of the transverse spin Seebeck effect in ferromagnetic conductors.

\section{Acknowledgements}

The authors gratefully acknowledge financial support by the Deutsche Forschungsgemeinschaft (DFG) within the priority programme SpinCat (RE 1052/24-1 and BA2181/11-1) and the European Metrology Research Programme EMRP (EURAMET) as well as the EU-ITN SpinIcur.

\section{References}

\bibliographystyle{apsrev4-1}
\bibliography{bibfile}

\end{document}